\pdfoutput=1

\documentclass[11pt,twoside,a4paper,cmspaper,final,collab]{cms-tdr}
\def\svnVersion{63579:63623}\def\cmsCernNo{2011-070}\def\cmsCernDate{\today}\def\cmsMessage{Submitted to Physical Review Letters}

\begin{document}\cmsNoteHeader{EXO-11-003}

\hyphenation{had-ron-i-za-tion}
\hyphenation{cal-or-i-me-ter}
\hyphenation{de-vices}
\RCS$Revision: 63623 $
\RCS$HeadURL: svn+ssh://alverson@svn.cern.ch/reps/tdr2/papers/EXO-11-003/trunk/EXO-11-003.tex $
\RCS$Id: EXO-11-003.tex 63623 2011-06-23 17:04:24Z alverson $
\providecommand{\cPg}{\ensuremath{\cmsSymbolFace{g}}} 
\providecommand{\cPG}{\ensuremath{\cmsSymbolFace{G}}} 
\providecommand{\cPU}{\ensuremath{\cmsSymbolFace{U}}} 
\providecommand{\cPZ}{\ensuremath{\cmsSymbolFace{Z}}} 
\newcommand{\LambU}{\ensuremath{{\Lambda_\cmsSymbolFace{U}}}\xspace}
\newcommand{\dU}{\ensuremath{{d_\cmsSymbolFace{U}}}\xspace}
\cmsNoteHeader{EXO-11-003}

\title{\texorpdfstring{Search for New Physics with a Mono-Jet and Missing Transverse Energy in $\Pp\Pp$ Collisions at $\sqrt{s}$ = 7~TeV}{Search for New Physics with a Mono-Jet and Missing Transverse Energy in pp Collisions at sqrt(s) = 7 TeV}}

\address[cern]{CERN}
\author[cern]{The CMS Collaboration}

\date{\today}

\abstract{
A study of events with missing transverse energy and an
energetic jet is performed using $\Pp\Pp$ collision data at a
centre-of-mass energy of 7\TeV.  The data were collected by
the CMS detector at the LHC, and
correspond to an integrated luminosity of 36\pbinv.
An excess of these events over standard model contributions
is a signature of new physics such as large extra dimensions
and unparticles.
The number of observed events is in good agreement with the
prediction of the standard model, and
significant extension of the current limits on parameters
of new physics benchmark models is achieved.}

\hypersetup{%
pdfauthor={CMS Collaboration},%
pdftitle={Search for New Physics with a Mono-Jet and Missing Transverse Energy in pp Collisions at sqrt(s) = 7 TeV},%
pdfsubject={CMS},%
pdfkeywords={CMS, physics, ADD, extradimensions, unparticle, monojet}}

\maketitle

This Letter describes a search for new physics
in the missing transverse energy (\MET) and jet final state using data
collected with the Compact Muon Solenoid (CMS) experiment in
$\Pp\Pp$ collisions at a centre-of-mass energy of 7\TeV provided
by the Large Hadron Collider (LHC).
Events containing a single energetic jet (mono-jet) are selected,
although a second less energetic jet is allowed.
This event signature is predicted in models such as
large extra dimensions, based on the scenario by Arkani-Hamed,
Dimopoulos, and Dvali (ADD) \cite{bib:INTRO_ADD1,bib:INTRO_ADD2,bib:INTRO_ADD3,bib:INTRO_ADD4}, or
unparticles~\cite{bib:INTRO_Unp}. The data used in this search
were collected in 2010 and correspond to an integrated luminosity of
36\pbinv.  This study focuses on the search for direct production of
a graviton $\cPG$ (or unparticle $\cPU$) balanced by a
hadronic jet via the processes $\Pq\Paq \rightarrow \cPg\cPG$ $(\cPg\cPU)$,
$\cPq\cPg \rightarrow \cPq\cPG$ $(\cPq\cPU)$ and $\cPg\cPg \rightarrow \cPg\cPG$ $(\cPg\cPU)$.
The primary backgrounds for this search are from $\cPZ$+jet and $\PW$+jet production, and are
estimated from the data.

The ADD model aims at explaining the large difference between
the electroweak and Planck scales by introducing a number $\delta$ of
extra spatial dimensions which in the simplest scenario are
compactified over a torus of common radius $R$.
The fundamental scale $\MD$
is related to the effective four-dimensional Planck scale $\Mpl$
according to the formula $\Mpl^2\approx \MD^{\delta+2} R^{\delta}$.
Gravitons are assumed to propagate in the extra dimensions with
nonzero momentum.  Graviton production is expected to be greatly
enhanced due to the kinematically available phase space in the extra
dimensions. Once produced, the gravitons are very weakly
coupled and their presence can only be inferred from \MET.
Searches in both the jet plus \MET and the
$\Pgg$ plus \MET channels were performed previously
\cite{bib:INTRO_ALEPH,bib:INTRO_OPAL,bib:INTRO_DELPHI,bib:INTRO_L3,bib:INTRO_CDF,bib:INTRO_D0},
without any
evidence of new physics observed.  The current lower limits
on $\MD$ range from 1.6\TeVcc for $\delta=2$ \cite{bib:INTRO_ALEPH,bib:INTRO_OPAL,bib:INTRO_DELPHI,bib:INTRO_L3} to
0.95\TeVcc for $\delta=6$~\cite{bib:INTRO_CDF}.

More recently, interest in unparticle models has increased.
These models relate to physics
originating from a new scale-invariant (conformal) sector, which
is coupled to the standard model (SM) through a connector sector at a high mass
scale.
An operator with a general non-integer scale dimension $\dU$ in a conformal sector
induces
a spectrum of invisible, massless, and weakly interacting particles.
If the mass scale $\LambU$ is assumed to be roughly 1~\TeVcc, then
by using an effective field theory below that scale one should be
able to study the effects of unparticles at the LHC.
While there have been no direct searches for unparticles, a recent
interpretation of CDF results suggests lower limits on $\LambU$
between 2.11 and 9.19\TeVcc for $1.05<\dU<1.35$ \cite{bib:INTRO_Delgado,bib:INTRO_Kathrein}.

The CMS apparatus has pixel and silicon-strip detectors for
pseudorapidity of $|\eta| < 2.5$, where $\eta=-\ln[\tan(\theta/2)]$
and $\theta$ is the polar angle relative to the beam direction.
Contained in
a 3.8~T magnetic solenoid, the tracking detectors
provide momentum reconstruction down to about 100\,\MeVc with a
resolution of about 1\% at 100\GeVc.
A highly granular crystal electromagnetic calorimeter (ECAL)
extends to $|\eta| < 3.0$, and has an energy resolution of
better than 0.5\% for photons with a transverse energy above
100\GeV.
A hermetic hadronic calorimeter (HCAL) extends to $|\eta| < 5.0$
with a transverse hadronic energy resolution of about
$100\,\%/\sqrt{E_{\rm T}\,[\GeV]} \oplus 5\,\%$.
A muon detector system reconstructs and identifies muons
to $|\eta| < 2.4$.
A full description of the CMS detector can be
found in Ref.~\cite{bib:CMS_TDR}.

Both ADD and unparticle signal events are
generated with the \PYTHIA 8.130 Monte Carlo
generator~\cite{bib:GEN_PY8,bib:GEN_Ask} with Tune 1 and passed through the CMS full simulation via the
\GEANTfour package~\cite{bib:GEN_GEANT4}.  The
CTEQ~6.6M parton distribution functions (PDFs)~\cite{bib:SYST_CTEQ6M}
are used throughout.  In order to scan the sensitivity in
the relevant ADD parameter space, different samples with
$\MD=1,2,3 \TeVcc$ and $\delta=2,3,4,5,6$ are produced.
The models are effective theories and hold only for energies
well below $\MD$ ($\LambU$),
we therefore follow the convention to suppress the simulated cross section of the graviton (unparticle) by a factor $\MD^2/\hat{s}$ ($\LambU^2/\hat{s}$) above $\MD$ ($\LambU$).
A transverse momentum (\pt) cutoff on the
parton recoiling against the graviton (unparticle) is introduced by
requiring $\hat{p}_{\rm T}>$50\GeVc at parton level, where
$\hat{p}_{\rm T}$ is the transverse momentum of the outgoing parton
(gluon or quark) from the initial hard scatter.
In this analysis, unparticles are assumed sufficiently long-lived
that they do not decay in the detector.
The next-to-leading-order (NLO) QCD corrections
to the direct graviton plus jet production in ADD are sizable and dependent
on the \pt of the recoiling parton~\cite{bib:Grav_NLO}.
$K$ factors ($\sigma_{\textrm{NLO}}/\sigma_{\textrm{LO}}$) are chosen
for a graviton transverse momentum of several hundred $\!\GeVc$, corresponding
to $1.5$ for $\delta=2,3$ and $1.4$ for $\delta=4,5,6$.
The background samples of vector boson plus jets and top quark
pairs are produced with \MADGRAPH~\cite{bib:GEN_Mg} and
simulated using \PYTHIA (v6.420)~\cite{bib:GEN_Py6} with tune
D6T~\cite{bib:GEN_Py6_D6T} for showering, based on a leading-order
(LO) calculation of the matrix element with the shower
matching prescription~\cite{bib:GEN_MLM},
and interfaced with \TAUOLA~\cite{bib:GEN_Tauola}.
The QCD multijet background sample is generated with \MADGRAPH and also
interfaced to \PYTHIA with tune D6T for showering~\cite{bib:GEN_Py6,bib:GEN_Py6_D6T}.

\label{sec:ANA}

Several jet and \MET triggers are used for data collection,
and all trigger paths are fully efficient for events with a value of
$\MET>$120\GeV reconstructed offline.
Events are required to have at least one primary vertex, where the
vertex is reconstructed within a 24 cm window along the beam axis,
has a transverse distance from the beam spot no more than 2 cm,
and be of good quality~\cite{bib:ANA_Tk}.
Beam halo and other beam-induced background events are rejected by
requiring at least 25\% of the tracks
in events with ten or more tracks
to be well reconstructed~\cite{bib:ANA_BeamHalo}.

Jets and \MET are reconstructed using a particle flow
technique~\cite{bib:ANA_PF}. The algorithm produces a
unique list of particles in each event, using the combined
information from all CMS subdetectors. This list is then used as
input to the jet clustering, which reconstructs jets using the
anti-$\kt$ algorithm~\cite{bib:ANA_AK} with a distance parameter
of 0.5.
The missing transverse energy vector is computed as the
negative vector sum of the transverse momenta
of all particles reconstructed in the event, and has a
magnitude denoted by \MET.

Jet energies
are corrected to establish a uniform calorimeter response in
$\eta$ and an absolute response in \pt calibrated
at the particle level.  Jet-energy-scale corrections are derived
from Monte Carlo simulation (MC), and a residual correction is
derived by measuring the \pt balance in dijet
events~\cite{bib:ANA_JME-10-010}.
Jets are required to have $\pt>30\GeVc$. 
To remove any artificial
signals in the calorimeter, criteria based on energy  sharing
between neighbouring channels are applied~\cite{bib:ANA_JME-10-004}.
Signals in  HCAL or ECAL towers
identified to be unphysical are removed from the reconstruction.
Beam halo and cosmic muons are removed, but some of these events
leave energy in both the ECAL and HCAL with no charged track associated
with the energy cluster in the calorimeter.
The fraction of jet energy carried by charged hadrons is therefore
required to be above 15\%.  To reject high-\pt photons
and electrons misidentified as hadronic jets, the
energies assigned to neutral hadrons in the HCAL and neutral and
charged hadrons in the ECAL must sum to less than 80\% of the
total jet energy.
The combined effect of all data cleanup is to reject 1.5\% of
the events in the signal sample defined below.

In order to reduce the background from $\PW$-boson decays, events
with isolated leptons are rejected.
A separate $\PW$ boson enriched sample is also created by requiring
isolated leptons, and is used to estimate the size of the primary
background.
Lepton candidates (electron and muon) are required to have
$\pt>20\GeVc$, to originate within 2\,mm of the beam axis in the
transverse plane, and to be spatially separated from jets by at
least $\DR=\sqrt{(\Delta \eta)^2+(\Delta \phi)^2}=0.5$
in order to avoid rejecting events where there are leptons from jets.
Here $\Delta\eta$ and $\Delta\phi$ are the pseudorapidity
and azimuthal angle (in radians) differences, respectively.
Muon candidates within $|\eta|<2.1$ are reconstructed by requiring
both that compatible tracks in the silicon tracking detectors are
found, and that these signals are consistent with a global fit to
both silicon tracker and muon detector hit locations~\cite{bib:ANA_muons}.
Electron candidates are reconstructed
starting from a cluster of energy deposits in the
ECAL, which is then matched to hits in the silicon tracker. Electron
candidates are required to have $|\eta|<1.44$ or
$1.56<|\eta|<2.5$ to avoid poorly-instrumented regions, and
candidates with significant mismeasurement in the ECAL or consistent
with a photon conversion are rejected~\cite{bib:ANA_electrons}.
For lepton candidates a cone of $\DR<0.3$ is constructed
around the track direction. 
An isolation parameter is defined as the scalar sum of the transverse momenta of
tracks and transverse energies in the ECAL and HCAL in the cone,
excluding the contribution from the muon (electron) candidate,
divided by the muon \pt (electron \ET).
Candidates
with isolation values below 0.15 for muons or 0.09 (0.04) for
electrons in the central (forward) regions are considered isolated.

\begin{figure}
  \begin{center}
  \includegraphics[scale=0.39]{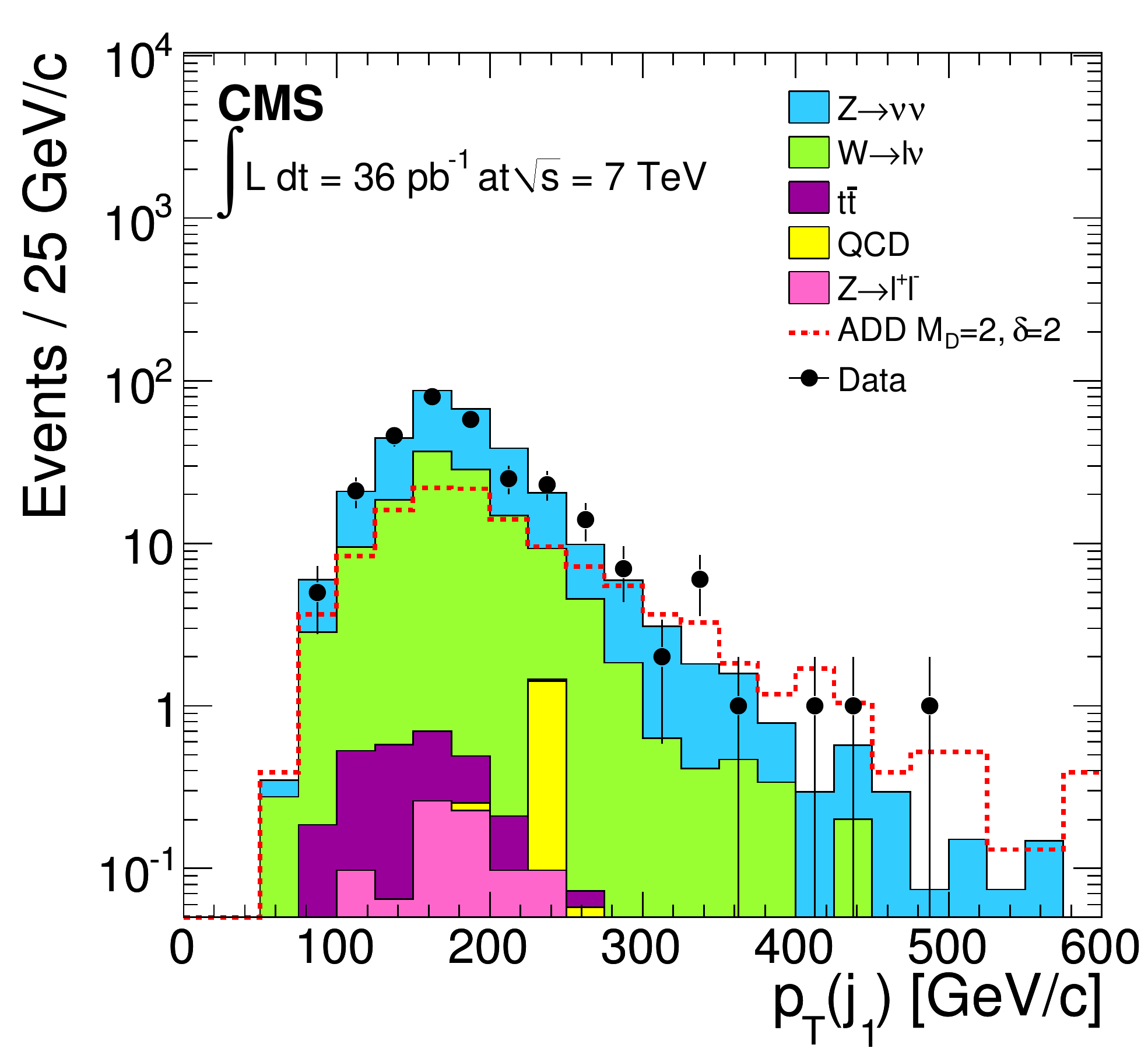}
  \caption{Distribution of $\pt(j_1)$, requiring
\MET $>$ 150\GeV,
 N$_\textrm{jets}\le 2$, $|\eta(j_1)|\!<\!2.4$, and $\Delta\phi(j_1,j_2)<2$.  A
representative ADD signal (with $\MD=2\TeVcc$, $\delta=2$) is shown as a dashed red line.
The background is normalised to the measured rate in data.
         \label{fig:ANA_Jet1_plot}}
  \end{center}
\end{figure}

The signal sample is selected by requiring $\MET>150\GeV$,
the most energetic jet ($j_1$) to have $\pt(j_1)>110\GeVc$ and
$|\eta(j_1)|<2.4$.
Events with more than two jets
(N$_\textrm{jets}> 2$)
with $\pt$ above 30\GeVc are discarded.
To increase the signal efficiency a second jet ($j_2$) is allowed provided
that the angular separation with the highest-\pt jet satisfies
$\Delta\phi(j_1,j_2)<2.0$ radians, a selection that suppresses
QCD dijet events.
The $\pt(j_1)$ distribution of the signal sample is shown in Fig.~\ref{fig:ANA_Jet1_plot}.
Remaining events with an isolated track are eliminated, as they
come primarily from $\Pgt$ decays.  A hollow
cone $0.02<\DR<0.3$ is defined around each track with $\pt > 10\GeVc$.
The scalar sum of the \pt of all tracks with $\pt > 1\GeVc$ inside
the cone is calculated and the event is vetoed if this sum is smaller
than 10\% of the \pt of the original track.

\begin{figure*}
  \begin{center}
  \includegraphics[scale=0.39]{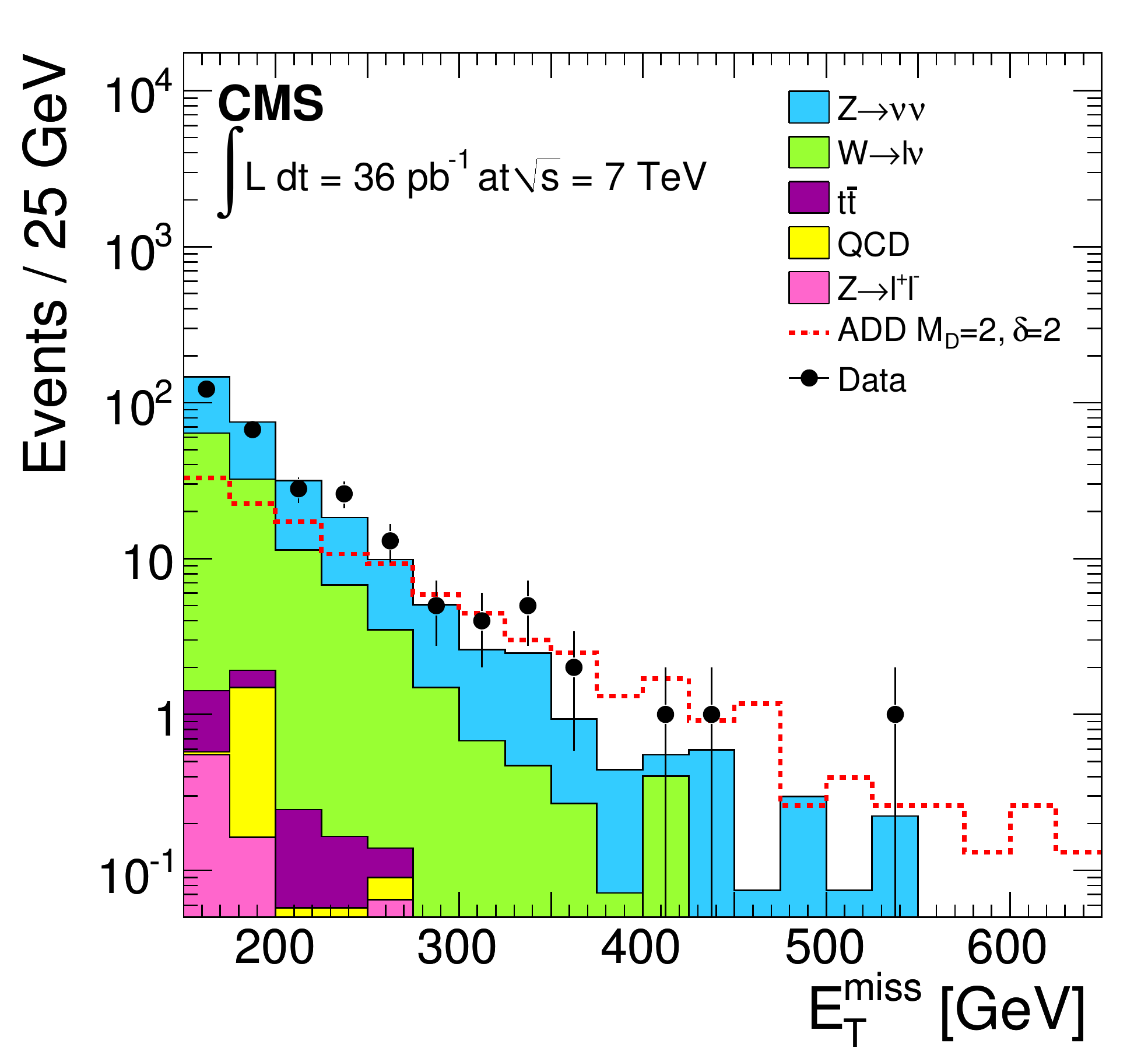}
  \includegraphics[scale=0.39]{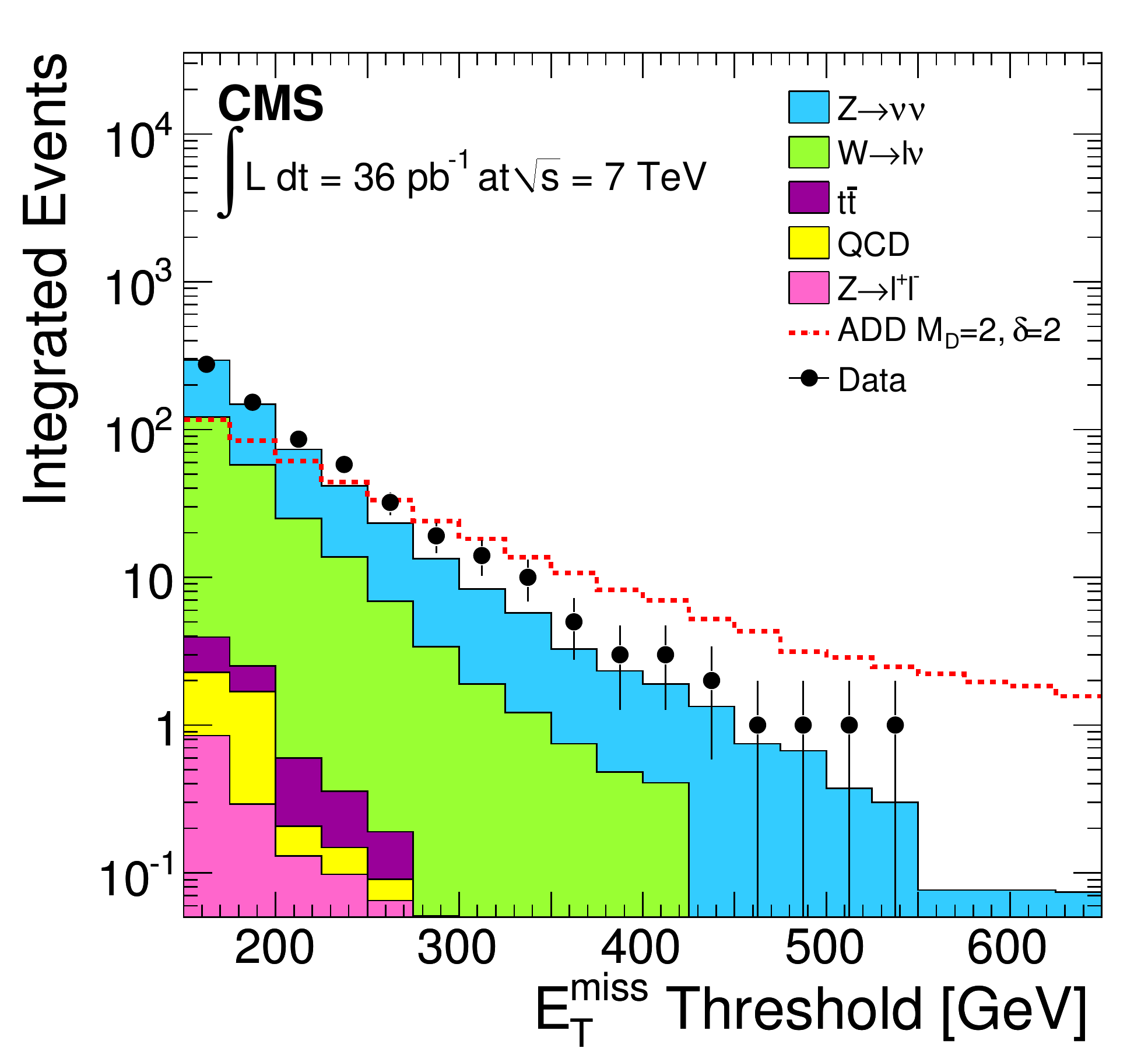}
  \caption{Missing transverse energy \MET after all selection
cuts for data, SM background, and an example of ADD signal
($\MD$=2\TeVcc, $\delta$=2). The figure at right
shows the integrated number of events above a given threshold.
The background is normalised to the measured rate in data.
         \label{fig:ANA_MET_plots}}
  \end{center}
\end{figure*}

The \MET distribution from data and the expected backgrounds after all
selection criteria are shown in Fig.~\ref{fig:ANA_MET_plots}, together with
a distribution of the integrated number of \MET events above a given threshold.
The only significant remaining backgrounds after all requirements stem
from electroweak processes with genuine missing transverse energy in
the final state.
Table~\ref{tab:SEL_TabDataMC} lists the number of events selected at
each step of the analysis from data and simulation.

\begin{table*}
        \begin{center}
        \caption{Mono-jet data sample and analysis cuts, with
luminosity-normalised leading-order MC.
Lepton removal eliminates  isolated muons
or tracks for $\pt (\Pe\textrm{, }\Pgm)\!>10\GeVc$.}
        \label{tab:SEL_TabDataMC}
                \begin{tabular}{lccccc|cc} \hline
Requirement & $\PW$+jets& $\cPZ(\cPgn\cPgn)$+jets& $\cPZ(\ell\ell)$+jets & \ttbar & QCD & Total MC & Data\\\hline
$\MET\!>\!150$\GeV, jet cleaning
                     & 622   &  259  & 46.7   &  90.4   & 202     & 1220 & 1298\\
$\pt\!(j_1)>\!110\GeVc$, $|\eta(j_1)|\!<\!2.4$
                     & 583   &  245  & 43.4   &  76.9   & 201     & 1149 & 1193\\
N$_\textrm{jets}\le 2$
                     & 446   &  201  & 34.3   &  11.3   & 74.3      & 767  & 778 \\
$\Delta\phi(j_1,j_2)<2$
                     & 370   &  182  & 29.5   &  9.1    & 6.3       & 597  & 596 \\
Lepton Removal
                     & 107   &  173  & 0.8    &  1.7    & 1.4       & 284  & 275 \\\hline
                \end{tabular}
\end{center}
\end{table*}

\label{sec:BKG}

Rather than using the background estimates from MC shown in
Table~\ref{tab:SEL_TabDataMC}, the
$\cPZ$+jets with $\cPZ\rightarrow\cPgn\cPgn$
(denoted $\cPZ(\cPgn\cPgn)$+jets)
and $\PW$+jets
backgrounds are estimated from $\Pgm$+jet events derived from the data sample.
The selection defining this control sample has the same initial
requirements as for the signal region, except that one or more muons
are explicitly required.  Well-reconstructed and isolated muons are
selected following the criteria outlined above.
To ensure a pure $\PW$+jets sample, a single isolated muon
is required to form, with the $\MET$, a transverse mass $M_{\rm T}$
between 50 and 100\GeVcc.  The transverse mass is defined
as $M_{\rm T}=\sqrt{2\pt\MET\left(1-\cos(\Delta\phi\right))}$, where
$\Delta\phi$ is the angle in the transverse plane between the
$\pt$ the $\MET$ vectors.  Within the $M_{\rm T}$ window there
are 113 single-muon events in the data, compared to 103 estimated from
MC (95.3 $\PW$+jets, 2.9 $\PW(\Pgt\cPgn)$+jets, 2.4 $\cPZ$+jets, 2.4
$\ttbar$, and 0.08 from QCD multijets).
The shape and yield of the muon distributions observed in the data
are consistent with the expectation from SM sources.
We estimate the number of $\PW$+jets background events to be $117\pm16$.
This estimate is obtained by scaling the surviving $\PW$+jets MC events in
the signal sample by the ratio of observed and predicted $\PW$+jets events
in the muon sample.

To estimate the number of $\cPZ(\cPgn\cPgn)$+jets background events,
the number of muon events in the $M_{\rm T}$ window is rescaled by several
factors, including: (i) the ratio between the $\PW(\Pgm\cPgn)$+jets and
$\cPZ(\cPgn\cPgn)$+jets production cross sections, obtained by combining
the branching fractions of the decays~\cite{bib:BKG_PDG}
($0.553\pm 0.021$), (ii) the reciprocal of the kinematic and geometric
acceptance of the simulated sample ($2.40\pm 0.12$),
(iii) the efficiency of the lepton veto in the signal region taken from
simulation ($0.95\pm 0.02$), (iv) the spectral shape differences in
$\PW$+jets and $\cPZ$+jets for $\pt(\PW,\cPZ)>150\GeVc$ ($1.33\pm0.03$), and (v)
the correction for contributions other than $\PW(\Pgm\cPgn)$+jets, extracted
from LO MC ($0.923\pm 0.071$).
All uncertainties include both statistical and systematic effects.
The number of
$\cPZ(\cPgn\cPgn)$+jets events
in the signal region predicted from
$\PW$+jets events
is
$176\pm30$,
which agrees with the MC.
A crosscheck is made using two opposite-sign muons from
$\cPZ(\cPgn\cPgn)$+jets, where the 13 events with an
invariant mass consistent with that of a $Z$ boson gives
a prediction of $162\pm45$ background events.
The estimated number of events from all background
sources is $297\pm45$.
The uncertainty includes both statistical and systematic sources,
with correlations taken into account.

The most important uncertainties related to
theoretical signal modeling and experimental mismeasurement are
(i) the jet energy scale, simulated by shifting the jet four-vectors by
an $\eta$- and $\pt$-dependent factor related to the response,
yielding a variation of 3--7\% (7.5--11.5\%) for the ADD (unparticle)
signal efficiency \cite{bib:ANA_JME-10-010},
(ii) the jet energy resolution, estimated from a $\Pgg$+jet sample and
resulting in a 0.3--2.2\% (0.6--2.9\%) uncertainty on the ADD (unparticle)
signal acceptance \cite{bib:SYST_JME-10-014},
(iii) uncertainties on the PDFs, evaluated using
a reweighting technique with the CTEQ6M parameterisation~\cite{bib:SYST_CTEQ6M} and
resulting in a systematic uncertainty of 1--2\% (3--7\%) for the
ADD (unparticle) signal, and
(iv) a 4\% uncertainty on the luminosity measurement~\cite{bib:SYST_lumi}.
The total systematic uncertainties range from 6\% to 13\%, with the jet energy
scale uncertainty being the dominant one.

To interpret the consistency of the observed number of events with
the background expectation in the context of a model, and also to
facilitate comparison with previous results, we set exclusion
limits for both the ADD model and the unparticle scenario.
The upper limit on the number of non-SM events consistent with the
measurements is set using a Bayesian method~\cite{bib:STAT_RooStats,bib:BKG_PDG}
with a flat signal prior.
A log-normal density function is assigned to the background estimate
with the uncertainty derived from data.  The total uncertainty incorporates
the individual uncertainties on each background process and takes correlations
into account.

\begin{figure}
  \begin{center}
  \includegraphics[scale=0.39]{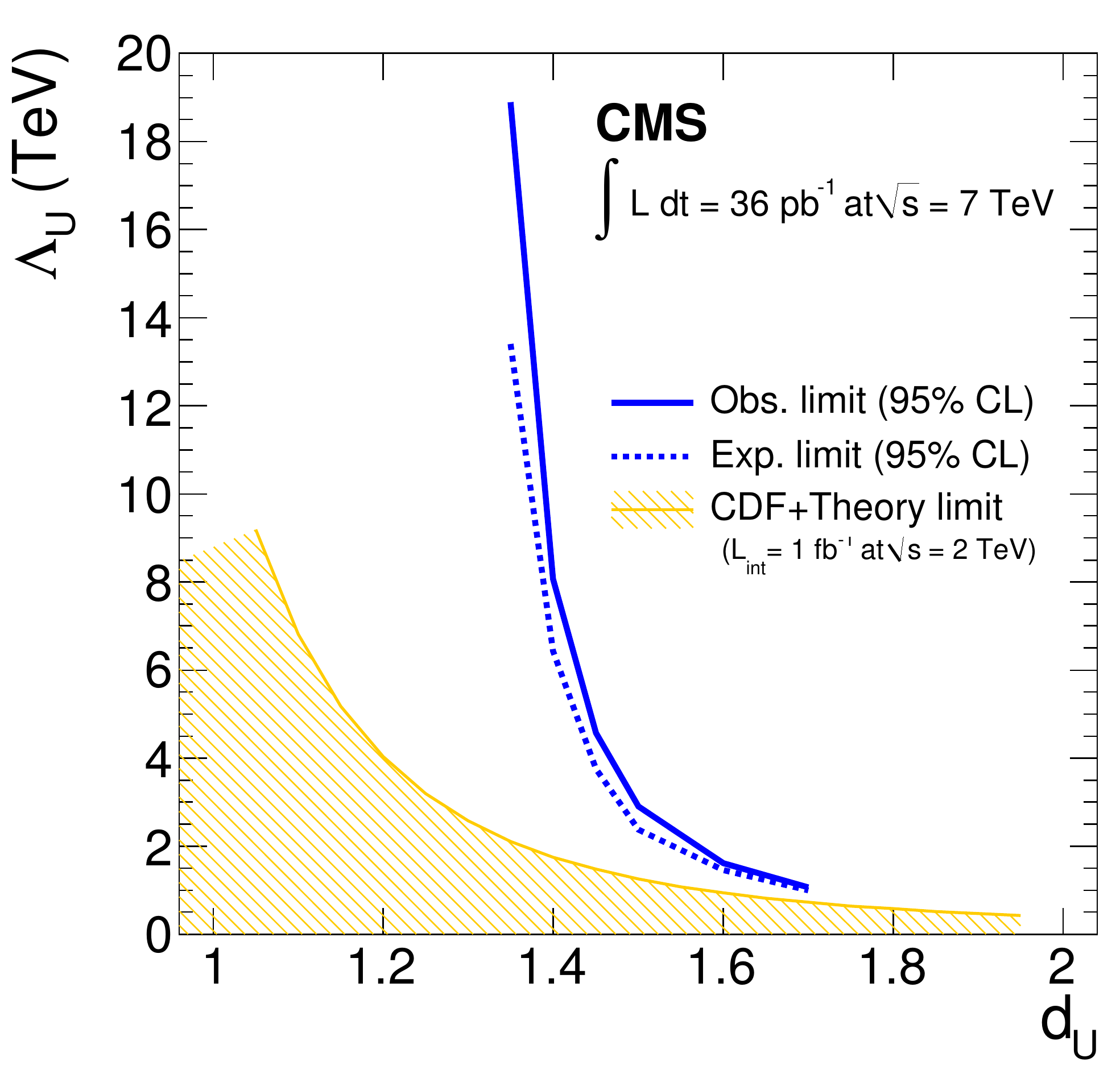}
  \caption{Observed and expected 95\% CL lower limits on the allowed region of unparticle model parameters $\dU$ and $\LambU$, compared to those derived from CDF results~\cite{bib:INTRO_Delgado,bib:INTRO_Kathrein}.}
\label{fig:STAT_Exclusion_Unp}
  \end{center}
\end{figure}

\begin{table}
        \begin{center}
        \caption{
Observed and expected 95\% CL lower limits on the ADD model parameter $\MD$
(in \TeVcc) as functions of $\delta$, with and without NLO
$K$ factors applied.
\label{tab:STAT_ADD_limits}}
        \begin{tabular}{c|c|cc|cc} \hline
~$\delta$~ & $K$ factor& LO Exp. & LO Obs. & NLO Exp. & NLO Obs. \\ \hline
 2  & 1.5 & 2.17 & 2.29 & 2.41 & 2.56  \\
 3  & 1.5 & 1.82 & 1.92 & 1.99 & 2.07  \\
 4  & 1.4 & 1.67 & 1.74 & 1.78 & 1.86  \\
 5  & 1.4 & 1.59 & 1.65 & 1.68 & 1.74  \\
 6  & 1.4 & 1.54 & 1.59 & 1.62 & 1.68  \\\hline
                \end{tabular}
\end{center}
\end{table}

Exclusion limits for the ADD model are given in Table~\ref{tab:STAT_ADD_limits} and
significantly improve the previous limits for this model.
For unparticles with spin = 0, production
cross sections above 54~pb are excluded at
95\% confidence level (CL) for $\dU=1.7$
and $\LambU=1\TeVcc$. The limits for other $\dU$ and $\LambU$ are
comparable and are shown in Fig.~\ref{fig:STAT_Exclusion_Unp};
for $\dU = (1.35, 1.40, 1.45, 1.50, 1.60, 1.70)$,
unparticles are excluded at 95\% CL for
$\LambU < (18.9, 8.07, 4.57, 2.90, 1.62, 1.07) \TeVcc$, compared to the expected limits of
$(13.4, 6.43, 3.75, 2.38, 1.46, 1.00) \TeVcc$.
From the ADD model with $\MD=3\TeVcc$ and $\delta=3$, which gives the largest
signal acceptance of 9.9\%, we evaluate a cross-section
upper limit for our selection
of 18.7~pb and exclude new processes at 95\% CL  above this
value that result in mono-jet events.

In summary, a search is performed for signatures from the ADD
and unparticle
models in events collected by the CMS experiment from
$pp$ collisions at $\sqrt{s}=7\TeV$.
A final state with an energetic jet
and a significant amount of missing transverse energy
is analyzed with the first CMS data, corresponding to
an integrated luminosity of 36\pbinv.
The QCD multijet background is reduced by several orders of
magnitude to a negligible level using topological cuts.
A measurement of the electroweak background from $\PW(\Pgm\cPgn)$-enriched
data is used to derive a background estimate for the
$\PW$+jets and $Z(\cPgn\cPgn)$+jets remaining in the signal region.  The data
are found to be in agreement with the expected contributions
from SM processes.  Limits on model parameters are derived
and constitute a significant improvement of those set by previous
experiments for ADD and unparticles.

We wish to congratulate our colleagues in the CERN accelerator departments for the excellent performance of the LHC machine. We thank the technical and administrative staff at CERN and other CMS institutes, and acknowledge support from: FMSR (Austria); FNRS and FWO (Belgium); CNPq, CAPES, FAPERJ, and FAPESP (Brazil); MES (Bulgaria); CERN; CAS, MoST, and NSFC (China); COLCIENCIAS (Colombia); MSES (Croatia); RPF (Cyprus); Academy of Sciences and NICPB (Estonia); Academy of Finland, ME, and HIP (Finland); CEA and CNRS/IN2P3 (France); BMBF, DFG, and HGF (Germany); GSRT (Greece); OTKA and NKTH (Hungary); DAE and DST (India); IPM (Iran); SFI (Ireland); INFN (Italy); NRF and WCU (Korea); LAS (Lithuania); CINVESTAV, CONACYT, SEP, and UASLP-FAI (Mexico); PAEC (Pakistan); SCSR (Poland); FCT (Portugal); JINR (Armenia, Belarus, Georgia, Ukraine, Uzbekistan); MST and MAE (Russia); MSTD (Serbia); MICINN and CPAN (Spain); Swiss Funding Agencies (Switzerland); NSC (Taipei); TUBITAK and TAEK (Turkey); STFC (United Kingdom); DOE and NSF (USA).

\bibliography{auto_generated}
\cleardoublepage\appendix\section{The CMS Collaboration \label{app:collab}}\begin{sloppypar}\hyphenpenalty=5000\widowpenalty=500\clubpenalty=5000\textbf{Yerevan Physics Institute,  Yerevan,  Armenia}\\*[0pt]
S.~Chatrchyan, V.~Khachatryan, A.M.~Sirunyan, A.~Tumasyan
\vskip\cmsinstskip
\textbf{Institut f\"{u}r Hochenergiephysik der OeAW,  Wien,  Austria}\\*[0pt]
W.~Adam, T.~Bergauer, M.~Dragicevic, J.~Er\"{o}, C.~Fabjan, M.~Friedl, R.~Fr\"{u}hwirth, V.M.~Ghete, J.~Hammer\cmsAuthorMark{1}, S.~H\"{a}nsel, M.~Hoch, N.~H\"{o}rmann, J.~Hrubec, M.~Jeitler, W.~Kiesenhofer, M.~Krammer, D.~Liko, I.~Mikulec, M.~Pernicka, H.~Rohringer, R.~Sch\"{o}fbeck, J.~Strauss, A.~Taurok, F.~Teischinger, P.~Wagner, W.~Waltenberger, G.~Walzel, E.~Widl, C.-E.~Wulz
% ...
\vskip\cmsinstskip
% ...
1:~~Also at CERN, European Organization for Nuclear Research, Geneva, Switzerland\\
% ...

\end{sloppypar}
\end{document}